\documentclass[aps,prd,showpacs,preprintnumbers,amsmath,amssymb]{revtex4}
\input epsf

\global\arraycolsep=2pt
\makeatletter
\def\fmslash{\@ifnextchar[{\fmsl@sh}{\fmsl@sh[0mu]}}
\def\fmsl@sh[#1]#2{%
  \mathchoice
    {\@fmsl@sh\displaystyle{#1}{#2}}%
    {\@fmsl@sh\textstyle{#1}{#2}}%
    {\@fmsl@sh\scriptstyle{#1}{#2}}%
    {\@fmsl@sh\scriptscriptstyle{#1}{#2}}}
\def\@fmsl@sh#1#2#3{\m@th\ooalign{$\hfil#1\mkern#2/\hfil$\crcr$#1#3$}}
\makeatother

\begin{document}
\title{Excited heavy baryon masses to order $\Lambda_{QCD}/m_Q$ from
 QCD sum rules}
\author{Dao-Wei Wang}
\affiliation{Department of Applied Physics, National University of Defense
Technology, Hunan 410073, China}
\author{Ming-Qiu Huang}
\affiliation{~CCAST (World Laboratory) P.O. Box 8730, Beijing 100080, China}
\affiliation{Department of Applied Physics, National University of Defense
Technology, Hunan 410073, China}
\date{\today}
\begin{abstract}
Masses of the p-wave excited heavy baryons have been calculated to
the $\Lambda_{QCD}/m_Q$ order using QCD sum rule method within the
framework of heavy quark effective theory. Numerical results for
kinetic energy $\lambda_1$ and chromo-magnetic interaction
$\lambda_2$ are presented. The splitting between spin $1/2$ and
$3/2$ doublet derived from our calculation is given, for which the
agreement with the current experiment is desirable.
\end{abstract}
 \maketitle
\section{Introduction}
\label{sec1}

In the past decade continuous progress has been made in the
investigation of excited heavy baryons. The lowest lying orbitally
excited charmed states $\Lambda_c(2593)$ and $\Lambda_c(2625)$
have been observed by several collaborations\cite{lc-exp}, the
excited states of $\Xi_c$ have also been reported
recently\cite{xc-exp}, and more data are expected in the near
future. From the theoretical prospect of view those data need to
be studied comprehensively. Furthermore, with the collection of
more experimental data for exited heavy baryon states it is useful
to make some theoretical predictions on their spectroscopies.

Heavy baryons containing a single heavy quark can be described exactly by
heavy quark effective theory (HQET) \cite{HQET,review,KKP} in the heavy quark
limit. This fact should be contributed to the spin-flavor symmetry of system
comprised of infinitely heavy quarks. HQET has been applied successfully to
learn about the properties of heavy mesons and baryons, including the
spectroscopy and weak decays. The mass formula for a spin symmetry doublet of
heavy baryons up to order $1/m_Q$ corrections can be written as
\begin{equation}
\label{mass} M=m_Q+\bar\Lambda-\frac{1}{2 m_Q}(\lambda_1+d_M\,\lambda_2)\;,
\end{equation}
where parameter $\bar\Lambda$ is the effective mass of the light degrees of
freedom in the $m_Q\to\infty$ limit, $\lambda_1$ and $\lambda_2$ are related
to the heavy quark kinetic energy and the chromomagnetic energy of HQET,
respectively
\begin{eqnarray}
\label{matri}
\lambda_1&=&\langle B(v)\mid \bar h_v\,(i D^\perp )^2\,h_v\mid B(v)\rangle\;,\nonumber\\
d_M\,\lambda_2&=&\langle B(v)\mid \bar h_v\,\sigma_{\mu\nu}
\frac{g_s}{2}\,G^{\mu\nu}\,h_v\mid B(v)\rangle\;.
\end{eqnarray}
The constant $d_M$ characterizes the spin-orbit interaction of the heavy
quark and the gluon field, it is zero for singlet and $1\,$, $-\frac{1}{2}\,$
for spin 1/2, spin 3/2 doublet, respectively. Thus the splitting of the spin
$1/2$ and $3/2$ doublets is
\begin{equation}
\label{splitting} M^2_{B_{Q}^{*}}-M^2_{B_{Q}}=\frac{3}{2}\;\lambda_2\;,
\end{equation}
where $B_{Q}$, $B_{Q}^{*}$ denote spin 1/2 and 3/2 doublet, respectively.

The heavy baryon mass parameters $\lambda_1$ and $\lambda_2$ play a most
significant role in our understanding of the spectroscopy \cite{FaNe} and
inclusive decay rates \cite{Bigi}. They must be estimated in some
nonperturbative approaches due to the asymptotic freedom property of QCD. A
viable approach is the QCD sum rules \cite{svzsum} formulated in the
framework of HQET \cite{hqetsum}. This method allows us to relate hadronic
observable to QCD parameters {\it via} the operator product expansion (OPE)
of the correlator. In the case of ground state heavy baryon, predictions on
the mass spectroscopy have been made to leading and next-to-leading order in
$\alpha_s$ \cite{Grozin,Groote} and to order $1/m_Q$ \cite{Dai,wang} using
QCD sum rule method. As to the excited baryon mass spectroscopy, only results
to leading order in $1/m_Q$ expansion have been obtained from QCD sum rule
\cite{CSH,zhu}. In \cite{wang} we have calculated the baryonic parameters
$\lambda_1$ and $\lambda_2$ for the ground state $\Lambda_Q$ and $\Sigma_Q$
baryons using QCD sum rules in the HQET. Employing the baryonic currents from
\cite{CSH} we now derive these parameters for excited $\Lambda$- and
$\Sigma$-type baryons following the same procedure.

The remainder of this paper is organized as follows. In Sec.
\ref{ssec1} we introduce the interpolating currents for excited
state heavy baryons and briefly present the two-point sum rules.
The direct Laplace sum rules analysis for the matrix elements is
presented in Sec. \ref{ssec2}. The Sec. \ref{sec3} is devoted to
numerical results and our conclusions. Some comments are also
available in Sec. \ref{sec3}.

\section{derivation of the sum rules}
\label{sec2}
\subsection{Heavy baryonic currents and two-point sum rules}
\label{ssec1}

In this work we adopt those currents constructed from
Bethe-Salpeter equation in Ref.\cite{CSH} as
\begin{subequations}\label{current}
\begin{eqnarray}
j_{\Sigma_{Qk1}}&=&\epsilon_{abc}(q^{T\,a}_1\,\tau\,C\gamma_5\,D_{t}^{\mu_1}
\,q^{b}_2)\Gamma^\prime h_{v}^{c}\;,\\
j_{\Lambda_{Qk0}}&=&\epsilon_{abc}(q^{T\,a}_1\,\tau\,C\gamma_t^\mu\,D_{t\mu}
\,q^{b}_2)\Gamma^\prime h_{v}^{c}\;,\\
j_{\Lambda_{Qk1}}&=&\epsilon_{abc}(q^{T\,a}_1\,\tau\,C\epsilon_{\mu_1\nu\sigma\rho}
\gamma_t^{\nu}\, v^\rho\,D_{t}^{\sigma}
\,q^{b}_2)\Gamma^\prime h_{v}^{c}\;,\\
j_{\Lambda_{QK1}}&=&\epsilon_{abc}(q^{T\,a}_1\,\tau\,C\gamma_5
\,q^{b}_2)\Gamma^\prime\,D_{t}^{\mu_1} h_{v}^{c}\;,\\
j_{\Sigma_{QK0}}&=&\epsilon_{abc}(q^{T\,a}_1\,\tau\,C\gamma_t^\mu
\,q^{b}_2)\Gamma^\prime\,D_{t\mu} h_{v}^{c}\;,\\
j_{\Sigma_{QK1}}&=&\epsilon_{abc}(q^{T\,a}_1\,\tau\,C\epsilon_{\mu_1\nu\sigma\rho}
\gamma^{\nu}_t\, v^\rho\,q^{b}_2)\Gamma^\prime\,D_{t}^{\sigma} h_{v}^{c}\;,
\end{eqnarray}
\end{subequations}
in which $C$ is the charge conjugation matrix, $\tau$ is the
flavor matrix which is antisymmetric for $\Lambda$-type baryon and
symmetric for $\Sigma$-type baryon, $\Gamma^\prime$s are some
gamma matrices, and a, b, c denote the color indices.
$\Gamma^\prime$ can be chosen co-variantly as
\begin{equation}
\Gamma'=\gamma_{t\mu_1}\gamma_5\,
\end{equation}
for $\Sigma_{Qk1},\Lambda_{Qk1},\Lambda_{QK1},\Sigma_{QK1}$
doublets' spin 1/2 baryon, and
\begin{equation}
\Gamma'=\Gamma_{\mu_1\rho_1}=-\frac{1}{3}(g_{t\mu_1\rho_1}+
\gamma_{t\mu_1}\gamma_{t\rho_1}),
\end{equation}
for $\Sigma_{Qk1},\Lambda_{Qk1},\Lambda_{QK1},\Sigma_{QK1}$
doublets' spin 3/2 baryon, in which $g_{t\mu_1\rho_1}$ and
$\gamma_{t\mu_1}$ are perpendicular to heavy quark velocity $v$,
defined as $g_{t\mu\nu}=g_{\mu\nu}-v_\mu
v_\nu,\gamma_{t\mu}=\gamma_{\mu}-\rlap/v v_\mu$. For the singlets
$\Lambda_{Qk0}$ and $\Sigma_{QK0}$ the $\Gamma'$ is simply unit
matrix $I$. Notations used here to describe excited state heavy
baryons are the same as those used in Ref. \cite{KKP,CSH}: k
denotes $l_k=1$ and $l_K=0$ whereas K denotes $l_k=0$ and $l_K=1$,
in which $l_k$ is the orbital angular momentum describes the
relative motion of the two light quarks and the orbital angular
momentum $l_K$ describes orbital motion of the center of mass of
the two light quarks relative to the heavy quark. Q denotes heavy
quark and the number in subscript is the total angular momentum of
the light diquark system. As that for ground state baryons, the
flavor configuration of $\Lambda$ type baryon is antisymmetric and
$\Sigma$ type is symmetric. Depending on the number of derivatives
or the form of $\Gamma^\prime$, interpolating currents can have
forms different from those listed above and may also be used in
applications\cite{CSH,zhu}.

In the following analysis we would use those currents to interpolate excited
heavy baryon states. At the leading order of the $1/m_Q$ expansion they do
not mix with each other even with the same quantum number, but to the
next-to-leading order the mixing of interpolating currents will appear. In
our subsequent calculations we would only use those currents and did not
consider the effect resulting from the mixing of interpolating currents for
references \cite{Grozin,mixing1,mixing2} have shown that the stability
criterion for QCD sum rule applications excludes the existence of
interpolating currents mixing or though there does exist the mixing the
numerical result will not change drastically compared with the case without
mixing.

The baryonic coupling constant in HQET are also needed in our
calculation, they are defined in form as follows
\begin{equation}
\langle 0\mid j \mid B(v)\rangle=F\,u,
\end{equation}
where $\mid B(v)\rangle$ denotes excited baryon state and $u$ can
be the ordinary spinor u or the Rarita-Schwinger spinor $u_\alpha$
in the HQET corresponding to spin 1/2 or spin 3/2 doublet,
respectively. Irrespective of an irrelevant constant factor in the
leading order, the coupling constants for spin 1/2 and spin 3/2
doublet are equivalent for their identical spin-parity of the
light degrees of freedom.

In order to determine the effective mass of the excited baryons,
we analyze the two-point correlator defined as
\begin{equation}
\label{twop} i \int d^4x e^{ik\cdot x}\langle 0\mid T\{j(x)\bar
j(0)\}\mid 0\rangle=\frac{1+\rlap/v}{2}Tr[\tau \tau^+]\Pi
(\omega),
\end{equation}
where $k$ is the residual momentum and $\omega=v\cdot k$. For large negative
value of $\omega$ $\Pi(\omega)$ can be expressed in terms of perturbative and
nonperturbative contributions. The nonperturbative effects can be accounted
for by including quark and gluon condensates ordered by increasing dimension,
which are the series of power corrections in the "small" $1/\omega$ variable.
The Borel transformation in the variable $\omega$ can help to improve the
convergence of these nonperturbative series.

With those interpolating currents listed in Eq. (\ref{current}) it is
straightforward to obtain the two-point sum rules:
\begin{subequations}\begin{eqnarray}
F^{2}_{\Lambda_{Qk0}}\;e^{-\bar\Lambda_{\Lambda_{Qk0}}/T}&=&\frac{18\,T^8}{\pi^4}\,
\delta_7(\omega_c/T)
+\frac{T^4}{4\,\pi^2}\,\langle\frac{\alpha_s}{\pi}G^2\rangle\
+\frac{m_0^2}{16}\,\langle \bar q q\rangle^2\;e^{-\frac{m_0^2}{8T^2}}\;,\\
F^{2}_{\Sigma_{Qk1}}\;e^{-\bar\Lambda_{\Sigma_{Qk1}}/T}&=&\frac{90\,T^8}{\pi^4}\,\delta_7(\omega_c/T)
-\frac{3T^4}{2^5\,\pi^2}\,\langle\frac{\alpha_s}{\pi}G^2\rangle+\frac{m_0^2}{16}\,\langle
\bar q q\rangle^2\;e^{-\frac{m_0^2}{8T^2}}\;,\\
F^{2}_{\Lambda_{Qk1}}\;e^{-\bar\Lambda_{\Lambda_{Qk1}}/T}&=&\frac{216\,T^8}{\pi^4}\,
\delta_7(\omega_c/T)
-\frac{T^4}{4\,\pi^2}\,\langle\frac{\alpha_s}{\pi}G^2\rangle\
+\frac{m_0^2}{8}\,\langle \bar q q\rangle^2\;e^{-\frac{m_0^2}{8T^2}}\;,\\
F^{2}_{\Lambda_{QK1}}\;e^{-\bar\Lambda_{\Lambda_{QK1}}/T}&=&\frac{216\,T^8}{\pi^4}\,\delta_7(\omega_c/T)
-\frac{T^4}{8\,\pi^2}\,\langle\frac{\alpha_s}{\pi}G^2\rangle+\frac{m_0^2}{4}\,\langle
\bar q q\rangle^2\;e^{-\frac{m_0^2}{8T^2}}\;,\\
F^{2}_{\Sigma_{QK1}}\;e^{-\bar\Lambda_{\Sigma_{QK1}}/T}&=&\frac{288\,T^8}{\pi^4}\,\delta_7(\omega_c/T)
-\frac{3T^4}{2\,\pi^2}\,\langle\frac{\alpha_s}{\pi}G^2\rangle+\frac{m_0^2}{2}\,\langle
\bar q q\rangle^2\;e^{-\frac{m_0^2}{8T^2}}\;,\\
F^{2}_{\Sigma_{QK0}}\;e^{-\bar\Lambda_{\Sigma_{QK0}}/T}&=&\frac{504\,T^8}{\pi^4}\,\delta_7(\omega_c/T)
-\frac{T^4}{8\,\pi^2}\,\langle\frac{\alpha_s}{\pi}G^2\rangle+\frac{m_0^2}{4}\,\langle
\bar q q\rangle^2\;e^{-\frac{m_0^2}{8T^2}}.
\end{eqnarray}\end{subequations}
In calculations we adopted the gaussian ansatz for the nonlocal
quark condensate to get the dimension 6 condensate contribution.
Dimension $D > 6$ condensates are not included. The functions
$\delta_n(\omega_c/T)$ arise from the continuum subtraction and
are defined in \cite{wang}.

\subsection{Sum rules for $\lambda_1$ and $\lambda_2$}\label{ssec2}
For the evaluation of the matrix elements $\lambda_1$ and $\lambda_2$ we
consider the three-point correlators  as follows
\begin{eqnarray}
i^2\int d^4x\int d^4y e^{i k\cdot x-i k'\cdot y}\langle 0\mid T\{j(x)\;\bar
h_v\,(i D^\perp )^2\,h_v(0)\;\bar j(y)\}\mid 0\rangle
&=&\frac{1+\rlap/v}{2}Tr[\tau \tau^+]\;T_K(\omega,\omega')\;,\nonumber\\
i^2\int d^4x\int d^4y e^{i k\cdot x-i k'\cdot y}\langle 0\mid T\{j(x)\;\bar
h_v\,\sigma_{\mu\nu} \frac{g_s}{2}\,G^{\mu\nu}\,h_v(0)\;\bar j(y)\}\mid
0\rangle &=&d_M\frac{1+\rlap/v}{2}Tr[\tau
\tau^+]\;T_S(\omega,\omega')\;,\label{ms}
\end{eqnarray}
where the coefficients $T_K(\omega,\omega')$ and
$T_S(\omega,\omega')$ are analytic functions in the ``off-shell
energies'' $\omega=v\cdot k$ and $\omega'=v\cdot k'$ with
discontinuities for positive values of these variables. Saturating
the three-point functions with complete set of baryon states, one
can isolate the part of interest, the contribution of the
lowest-lying baryon states associated with the heavy-light
currents, as one having poles in both the variables $\omega$ and
$\omega'$ at the value $\omega=\omega'=\bar\Lambda$.

Confining us to take into account these leading contributions of
perturbation and the operators with dimension $D\leq 6$ in OPE,
the relevant diagrams in our theoretical calculations for the
kinetic energy are shown in Fig. 1. The relevant diagrams for the
chromo-magnetic interaction do not differ from those for the
ground state baryons \cite{wang}, so we do not show them here.
Using dispersion relations $T_K(\omega,\omega')$ and
$T_S(\omega,\omega')$ can be casted into the form of integrals of
the double spectral densities. Following
Refs.~\cite{IWfun,DHHL,Neubert92D}, introduce new variables
$\omega_+=\frac 12(\omega+\omega')$ and $\omega_-=\omega-\omega'$,
perform the integral over $\omega_-$, assume quark--hadron duality
in $\omega_+$, and employ Borel transformation $B^{\omega}_\tau$,
$B^{\omega^{\prime}}_{\tau^{\prime}}$ to suppress the continuum
contributions and subtractions, we then obtained the desired sum
rules. Considered the symmetry of the correlator it is natural to
set the parameters $\tau$, $\tau^{\prime}$ to be the same and
equal to $2T$, where $T$ is the Borel parameter of the two-point
functions. We ended up with the set of sum rules
\begin{subequations}\begin{eqnarray}
\lambda_{2}^{\Sigma_{QK1}}F^2e^{-\bar\Lambda/T}&=&\frac{2^{11}\;3}{\pi^4}
\frac{\alpha_s}{\pi}T^{10}\delta_9(\omega_c/T)
-\frac{2T^6}{\pi^2}\langle\frac{\alpha_s}{\pi}G^2\rangle
+\frac{16m_0^2\;T^2\alpha_s}{3\pi}\,\langle\bar q
q\rangle^2\;e^{-\frac{m_0^2}{16T^2}}\;,\\
\lambda_{2}^{\Lambda_{QK1}}F^2e^{-\bar\Lambda/T}&=&\frac{2^{10}\;3}{\pi^4}
\frac{\alpha_s}{\pi}T^{10}\delta_9(\omega_c/T)
-\frac{3T^6}{\pi^2}\langle\frac{\alpha_s}{\pi}G^2\rangle
+\frac{4m_0^2\;T^2\alpha_s}{3\pi}\,\langle\bar q
q\rangle^2\;e^{-\frac{m_0^2}{16T^2}}\;,\\
\lambda_{2}^{\Sigma_{Qk1}}F^2e^{-\bar\Lambda/T}&=&\frac{2^8\;3\;5}{\pi^4}
\frac{\alpha_s}{\pi}T^{10}\delta_9(\omega_c/T)
-\frac{T^6}{2\pi^2}\langle\frac{\alpha_s}{\pi}G^2\rangle
+\frac{m_0^2\;T^2\alpha_s}{3\pi}\,\langle\bar q
q\rangle^2\;e^{-\frac{m_0^2}{16T^2}}\;,\\
\lambda_{2}^{\Lambda_{Qk1}}F^2e^{-\bar\Lambda/T}&=&\frac{2^9\;3\;5}{\pi^4}
\frac{\alpha_s}{\pi}T^{10}\delta_9(\omega_c/T)
-\frac{3T^6}{\pi^2}\langle\frac{\alpha_s}{\pi}G^2\rangle
+\frac{4m_0^2\;T^2\alpha_s}{3\pi}\,\langle\bar q
q\rangle^2\;e^{-\frac{m_0^2}{16T^2}}\;,\\
-\lambda_{1}^{\Lambda_{QK1}}F^2e^{-\bar\Lambda/T}&=&
\frac{2^6\;3^3\;5T^{10}}{\pi^4}\delta_9(\omega_c/T)
+\frac{19\;T^6}{2\;\pi^2}\langle\frac{\alpha_s}{\pi}G^2\rangle
+\frac{5m_0^4}{16}\,\langle\bar q q\rangle^2\;e^{-\frac{m_0^2}{8T^2}}\;,\\
-\lambda_{1}^{\Sigma_{QK0}}F^2e^{-\bar\Lambda/T}&=&
\frac{2^6\;3^4\;5T^{10}}{\pi^4}\delta_9(\omega_c/T)
+\frac{113\;T^6}{4\;\pi^2}\langle\frac{\alpha_s}{\pi}G^2\rangle
+\frac{5m_0^4}{16}\,\langle\bar q q\rangle^2\;e^{-\frac{m_0^2}{8T^2}}\;,\\
-\lambda_{1}^{\Sigma_{QK1}}F^2e^{-\bar\Lambda/T}&=&
\frac{2^8\;3^2\;5T^{10}}{\pi^4}\delta_9(\omega_c/T)
-\frac{3\;T^6}{\pi^2}\langle\frac{\alpha_s}{\pi}G^2\rangle
+\frac{5m_0^4}{8}\,\langle\bar q q\rangle^2\;e^{-\frac{m_0^2}{8T^2}}\;,\\
-\lambda_{1}^{\Sigma_{Qk1}}F^2e^{-\bar\Lambda/T}&=&
\frac{2^4\;3^2\;17T^{10}}{\pi^4}\delta_9(\omega_c/T)
+\frac{21\;T^6}{8\;\pi^2}\langle\frac{\alpha_s}{\pi}G^2\rangle
+\frac{3m_0^4}{64}\,\langle\bar q q\rangle^2\;e^{-\frac{m_0^2}{8T^2}}\;,\\
-\lambda_{1}^{\Lambda_{Qk0}}F^2e^{-\bar\Lambda/T}&=&
\frac{2^4\;3^2\;7T^{10}}{\pi^4}\delta_9(\omega_c/T)
-\frac{17\;T^6}{8\;\pi^2}\langle\frac{\alpha_s}{\pi}G^2\rangle
+\frac{3m_0^4}{64}\,\langle\bar q q\rangle^2\;e^{-\frac{m_0^2}{8T^2}}\;,\\
-\lambda_{1}^{\Lambda_{Qk1}}F^2e^{-\bar\Lambda/T}&=&
\frac{2^6\;3^2\;11T^{10}}{\pi^4}\delta_9(\omega_c/T)
-\frac{5\;T^6}{2\;\pi^2}\langle\frac{\alpha_s}{\pi}G^2\rangle
+\frac{3m_0^4}{32}\,\langle\bar q q\rangle^2\;e^{-\frac{m_0^2}{8T^2}}\;.
\end{eqnarray}\end{subequations}
The unitary normalization of flavor matrix $Tr[\tau\tau^+]=1$ has
been applied to get those sum rules, as what has been done for the
two-point sum rules.

\section{numerical results and  conclusions}
\label{sec3}

In the following analysis, the standard value for the condensates are adopted
\cite{svzsum}. From two-point sum rules the effective mass can be obtained
via a derivative of the Borel parameter as $\bar\Lambda=T^2d\ln E/dT$, where
$E$ denotes the right hand side of the obtained two-point sum rule. Then
comply to the standard procedure of sum rule analysis, we changed the
continuum threshold $\omega_c$ and Borel parameter $T$ to find the optimal
stability window and the numerical value of the effective mass will be
determined within this window. For those sum rules obtained above, we found
the typical value for the continuum threshold is $\omega_c\sim 1.6\;
\mbox{GeV}$ and typical interval for the Borel parameter is $\Delta T\sim
0.4\; \mbox{GeV}$, which is a narrower one than the window for the ground
state baryon. The only exception of this assertion is $\Lambda_{Qk0}$, for
which we found a much lower continuum threshold $\omega_c\sim 1.2\;
\mbox{GeV}$ and a narrower interval $\Delta T\sim 0.3\; \mbox{GeV}$. Also it
is worth noting that for $\Lambda_{Qk0}$, the main contribution does not come
from the perturbative part. The condensate ones play an important role in the
determination of stability window. Indeed, if one keeps only gluon condensate
contribution, there will be no stability window at all, just like the case in
Ref.\cite{CSH}. But if assuming the perturbative dominance and omitting
condensate contributions, we only ended up with a better stability and the
numerical value is almost exactly the same. For the other sum rules for the
effective mass the case is different. The dominant contribution to those sum
rules comes from the perturbative part, and the dimension 4 and dimension 6
operators in the OPE only amount to 20\% and 10\% within the stability
windows, respectively. The numerical results are presented in Fig. 3. For the
aim of clarity we give our numerical average for the effective masses in
Table \ref{tab1}.

In order to get the numerical results for two $1/m_Q$ order parameters, we
divide our three-point sum rules by two-point functions to obtain $\lambda_1$
and $\lambda_2$ as functions of the continuum threshold $\omega_c$ and the
Borel parameter $T$. This procedure can eliminate the systematic
uncertainties and cancel the parameter $\bar\Lambda$. From the experiences of
QCD sum rule applications in the field of heavy quark physics it is well
known that three-point sum rule receives heavier contamination from the
continuum modes than two-point one, and the stability is not as good as that
for two-point sum rule\cite{BaBr,Grozin,wang,mixing2}. Our results for
$\lambda_1$ sum rules of those excited heavy baryons are shown in Fig. 3 and
in Fig. 4 results for $\lambda_2$ sum rules are presented. The typical value
of continuum threshold is $\omega_c\sim 1.8\; \mbox{GeV}$, $\omega_c\sim
2.4\;\mbox{GeV}$ for $\lambda_1$ and $\lambda_2$ sum rules, respectively. And
the typical interval of the stability window is also $\Delta T\sim
0.3\;\mbox{GeV}$ for both. The $\lambda_1$ sum rule for the $\Sigma_{QK0}$
baryon is an exception of above statement, for which the typical value of
continuum threshold is $\omega_c\sim 1.4\;\mbox{GeV}$ and the interval is
$\Delta T\sim 0.2\;\mbox{GeV}$. As for the convergence of the OPE within the
stability windows for those sum rules we would like to state some facts. For
the $\lambda_1$ sum rules, except the case of $\Lambda_{Qk1}$, in which the
dimension 4 and 6 operators give rise to contributions almost equal to that
of the perturbative part, all other sum rules behave well, in which the
contributions of the dimension 4 and 6 operators amount to 50\% and 20\% of
that of the perturbative part. On the other hand, for the $\lambda_2$ sum
rules, the dimension 4 operator still contributes to almost 50\% compared to
that of the perturbative part, but the dimension 6 operator gives only a
negligible contribution, typically less than 10\%. The numerical results are
listed in Table \ref{tab1}, too.

\begin{table}[htb]
\begin{tabular}{*{7}{@{\hspace{0.2cm}}c}}\hline\hline
&$\Lambda_{Qk0}$&$\Sigma_{Qk1}$&$\Lambda_{Qk1}$&$\Lambda_{QK1}$&$\Sigma_{QK1}$&$\Sigma_{QK0}$
\\\hline\hline
$\bar\Lambda$
&$1.04\pm0.17$&$1.12\pm0.22$&$1.36\pm0.18$&$1.30\pm0.13$&$1.27\pm0.11$&$1.21\pm0.09$
\\\hline
$-\lambda_1$
&$1.20\pm0.26$&$1.02\pm0.25$&$0.84\pm0.15$&$1.16\pm0.11$&$1.48\pm0.18$&$1.42\pm0.25$\\\hline
$\lambda_2$
&&$0.13\pm0.03$&$0.11\pm0.03$&$0.09\pm0.01$&$0.13\pm0.01$&\\\hline\hline
\end{tabular}
\caption{ Effective mass in GeV, kinetic energy and chromo-magnetic
interaction energy in GeV$^2$ for excited heavy baryons. Errors quoted are
due to the variation of the Borel parameter $T$ and continuum threshold
$\omega_c$.} \label{tab1}
\end{table}
With those values and heavy quark masses given in \cite{wang},
$m_c=1.41\;\mbox{GeV}$ and $m_b=4.77\;\mbox{GeV}$, masses of
excited heavy baryons to order $1/m_Q$ can be obtained
immediately. We give those masses in Table \ref{tab2}. Our results
are comparable to the prediction on the excited heavy baryon
masses obtained by using quark potential model \cite{mass}.

\begin{table}[htb]
\begin{tabular}{*{11}{@{\hspace{0.2cm}}c}}\hline\hline
&$\Lambda_{Qk0}$&$\Lambda_{Qk1}$&$\Lambda_{Qk1}^*$&$\Sigma_{Qk1}$&$\Sigma_{Qk1}^*$
&$\Lambda_{QK1}$&$\Lambda_{QK1}^*$&$\Sigma_{QK1}$&$\Sigma_{QK1}^*$&$\Sigma_{QK0}$
\\\hline\hline
Q=c&2.863&3.019&3.076&2.831&2.900&3.083&3.129&3.148&3.219&3.113\\\hline
Q=b&5.934&6.205&6.222&5.977&5.998&6.185&6.199&6.182&6.203&6.127\\\hline
\hline
\end{tabular}
\caption{ Masses in GeV for excited heavy baryons.} \label{tab2}
\end{table}

The splitting between spin 1/2 and spin 3/2 doublet can be
obtained by multiplying $\lambda_2$ by a factor $3/2$, cf. Eq.
(\ref{splitting}). It is $0.13\pm 0.02\;\mbox{GeV}^2$, $0.16\pm
0.05\;\mbox{GeV}^2$, $0.20\pm 0.02\;\mbox{GeV}^2$ and $0.19\pm
0.04 \;\mbox{GeV}^2$ for $\Lambda_{QK1}$, $\Lambda_{Qk1}$,
$\Sigma_{QK1}$ and $\Sigma_{Qk1}$ doublets, respectively. The
approximately equal value for $\Lambda_{QK1}$ and $\Lambda_{Qk1}$,
$\Sigma_{QK1}$ and $\Sigma_{Qk1}$ may be interpreted as the signal
which implies that current mixing effect cannot be large. If
taking the middle value as theoretical predictions for the
physical state, then the splitting for excited baryon states are
\begin{eqnarray}
\Lambda_{Q1}^{*\,2}-\Lambda_{Q1}^2&=&0.15\pm
0.03\;\mbox{GeV}^2,\nonumber\\
\Sigma_{Q1}^{*\,2}-\Sigma_{Q1}^2&=&0.20\pm 0.03\;\mbox{GeV}^2.
\end{eqnarray}
Based on the current experimental data\cite{data}, the splitting
of excited $\Lambda_c$ doublet is $0.17\;\mbox{GeV}^2$, which is
in agreement with our theoretical result. When it is scaled up to
the bottom quark mass scale there will be a factor $\sim 0.8$
approximately due to the renormalization group improvement.

To conclude, we have calculated the $1/m_Q$ order corrections to the excited
heavy baryon masses from QCD sum rules within the framework of the HQET. From
thus obtained spectrum for the c quark case, we found that $\Lambda_{Qk0}$
and $\Sigma_{Qk1}$ baryons lie $\sim600\;\mbox{MeV}$ above the ground state
baryon $\Lambda_c$, while $\Lambda_{Qk1}$, $\Lambda_{QK1}$, $\Sigma_{QK1}$
lie $\sim800\;\mbox{MeV}$ above $\Lambda_c$ and for $\Sigma_{QK1}$ baryon
this value is $\sim900\;\mbox{MeV}$, typically with an error $\sim
300\;\mbox{MeV}$. When it comes to the b quark case, the result is that
$\Lambda_{Qk0}$ and $\Sigma_{Qk1}$ baryons lie $\sim300\;\mbox{MeV}$ above
the ground state baryon $\Lambda_b$, while $\Lambda_{Qk1}$, $\Lambda_{QK1}$,
$\Sigma_{QK1}$ and $\Sigma_{QK1}$ lie $\sim550\;\mbox{MeV}$ above
$\Lambda_b$, for which the typical error is $\sim 200\;\mbox{MeV}$. For the c
quark case, $1/m_Q$ order corrections are $\sim 400\;\mbox{MeV}$, which is
not a small one due to the large value of the kinetic energy. When it comes
to the b quark case those corrections will be suppressed by the still larger
b quark mass. Our theoretical predictions for the doublet splitting are in
agreement with the current experimental data.
\acknowledgments We would like to thank Y.B. Dai for helpful discussions.
This work was supported in part by the National Natural Science Foundation of
China under Contract No. 10275091. \vspace{0.5cm}


\newpage
{\bf Figure Captions}

\begin{center}
\begin{minipage}{12cm}
{\sf Fig. 1.}{\quad Non-vanishing diagrams for the kinetic energy. The
kinetic energy operator is denoted by a white square, the interpolating
baryon currents by black circles. Heavy-quark propagators are drawn as double
lines.}
\end{minipage}
\end{center}

\begin{center}
\begin{minipage}{120mm}
{\sf Fig. 2.}{\quad Sum rules of the effective mass $\bar\Lambda$
for: (a) $\Lambda_{Qk0}$, (b) $\Lambda_{Qk1}$, (c) $\Sigma_{Qk1}$,
(d) $\Lambda_{QK1}$, (e) $\Sigma_{QK1}$ and (f) $\Sigma_{QK0}$
baryons. The different choices of the continuum threshold
$\omega_c$ corresponding to different curves are designated in
individual figures respectively. Curves are plotted against the
Borel parameter $T$.}
\end{minipage}
\end{center}

\begin{center}
\begin{minipage}{12cm}
{\sf Fig. 3.}{\quad Sum rules of the kinetic energy for: (a)
$\Lambda_{Qk0}$, (b) $\Lambda_{Qk1}$, (c) $\Sigma_{Qk1}$, (d)
$\Lambda_{QK1}$, (e) $\Sigma_{QK1}$ and (f) $\Sigma_{QK0}$
baryons. Others are the same as those in Fig. 2.}
\end{minipage}
\end{center}

\begin{center}
\begin{minipage}{12cm}
{\sf Fig. 4.}{\quad Sum rules of the chromo-magnetic interaction
for: (a) $\Lambda_{Qk1}$, (b) $\Sigma_{Qk1}$, (c) $\Lambda_{QK1}$
and (d) $\Sigma_{QK1}$ baryons. Others are the same as those in
Fig. 2.}
\end{minipage}
\end{center}

\newpage
\begin{figure} \vfill \epsfxsize=15cm
\begin{minipage}{.6\textwidth}
\epsfxsize=1.\textwidth \centerline{\epsffile{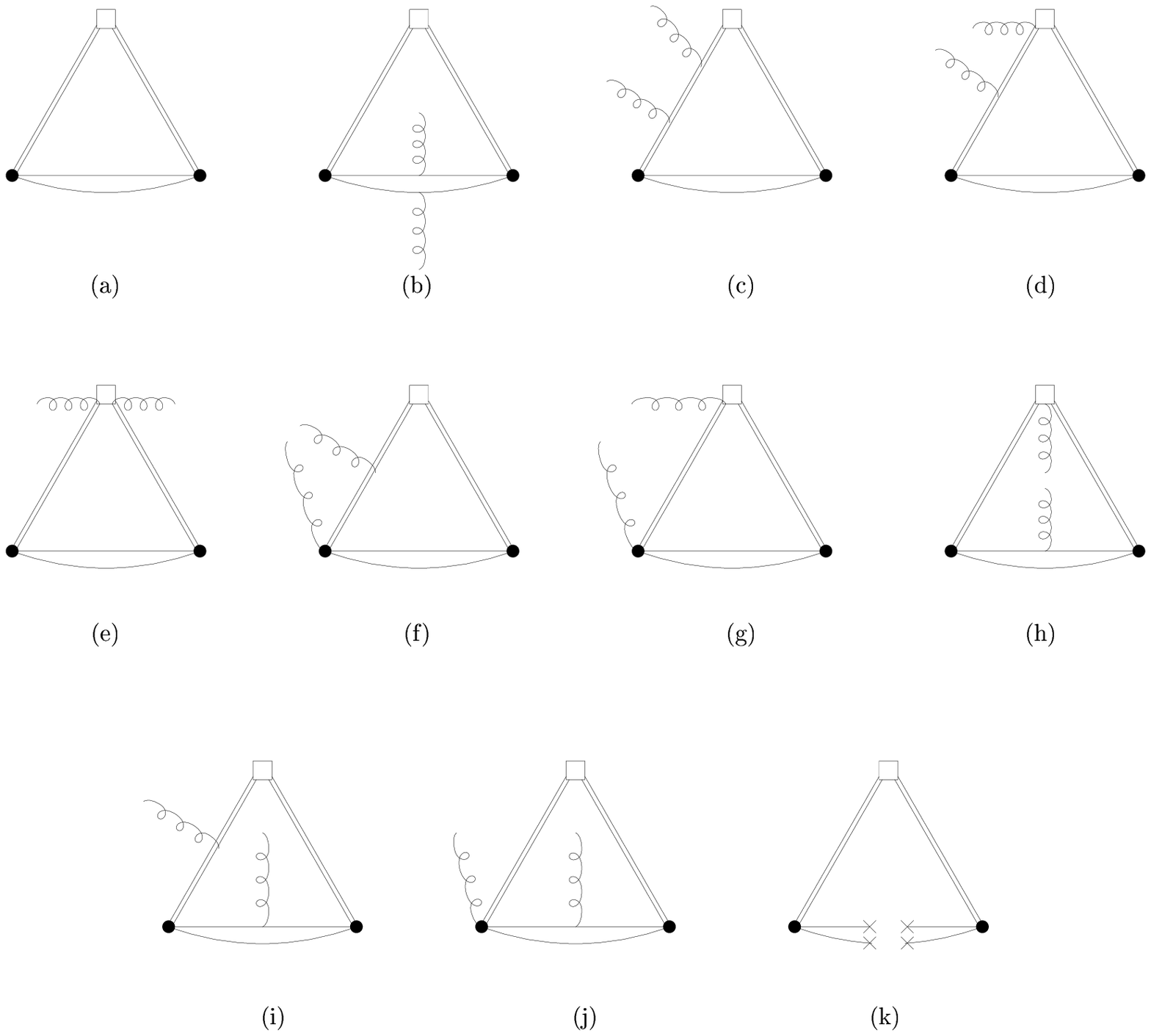}}
\end{minipage}
\center\vfill{\mbox{Fig. 1.}}
\end{figure}

\begin{figure}
\epsfxsize=15cm \centerline{\epsffile{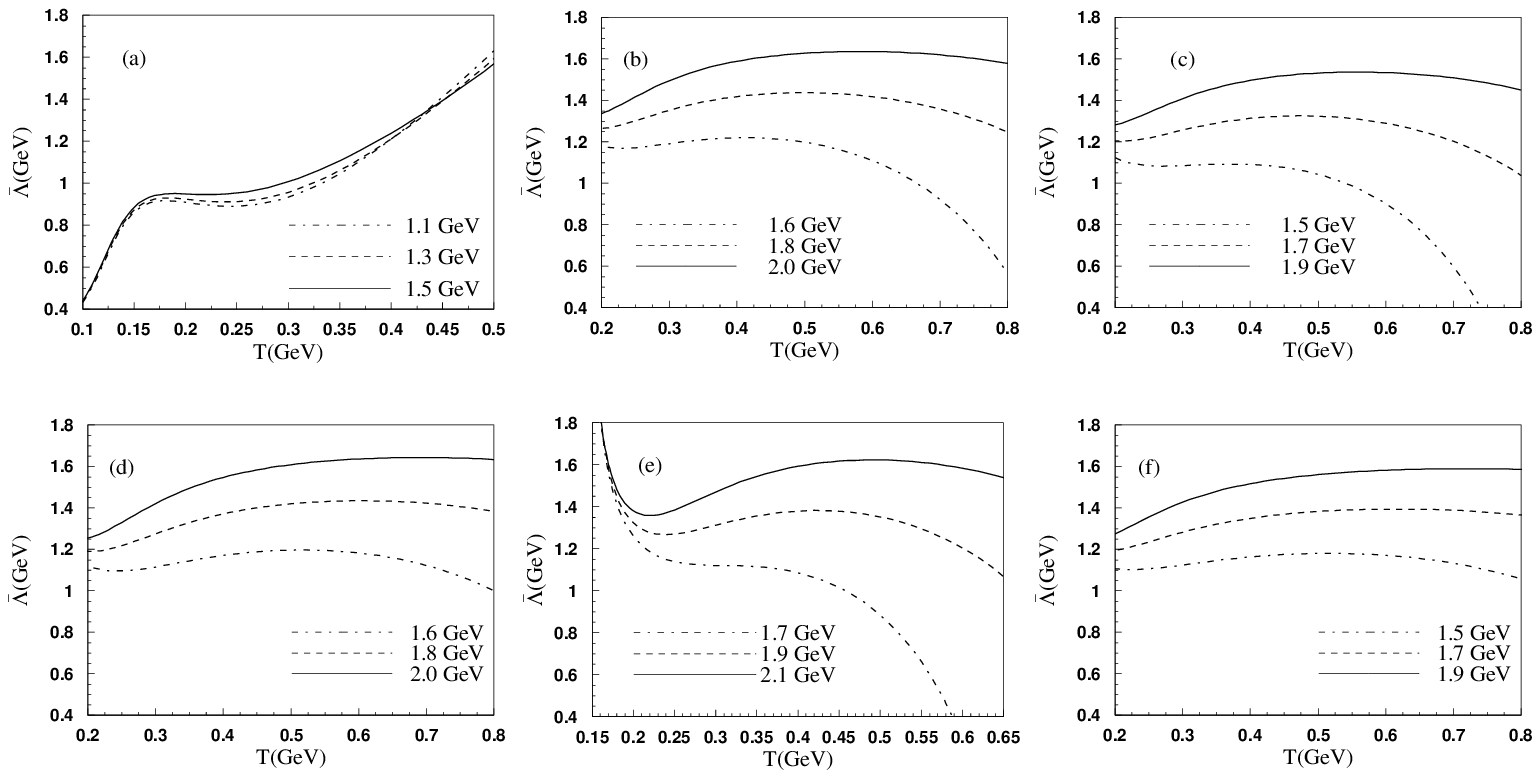}} \vfill \center{\mbox{Fig.
2.}}
\end{figure}

\begin{figure}
\epsfxsize=15cm \centerline{\epsffile{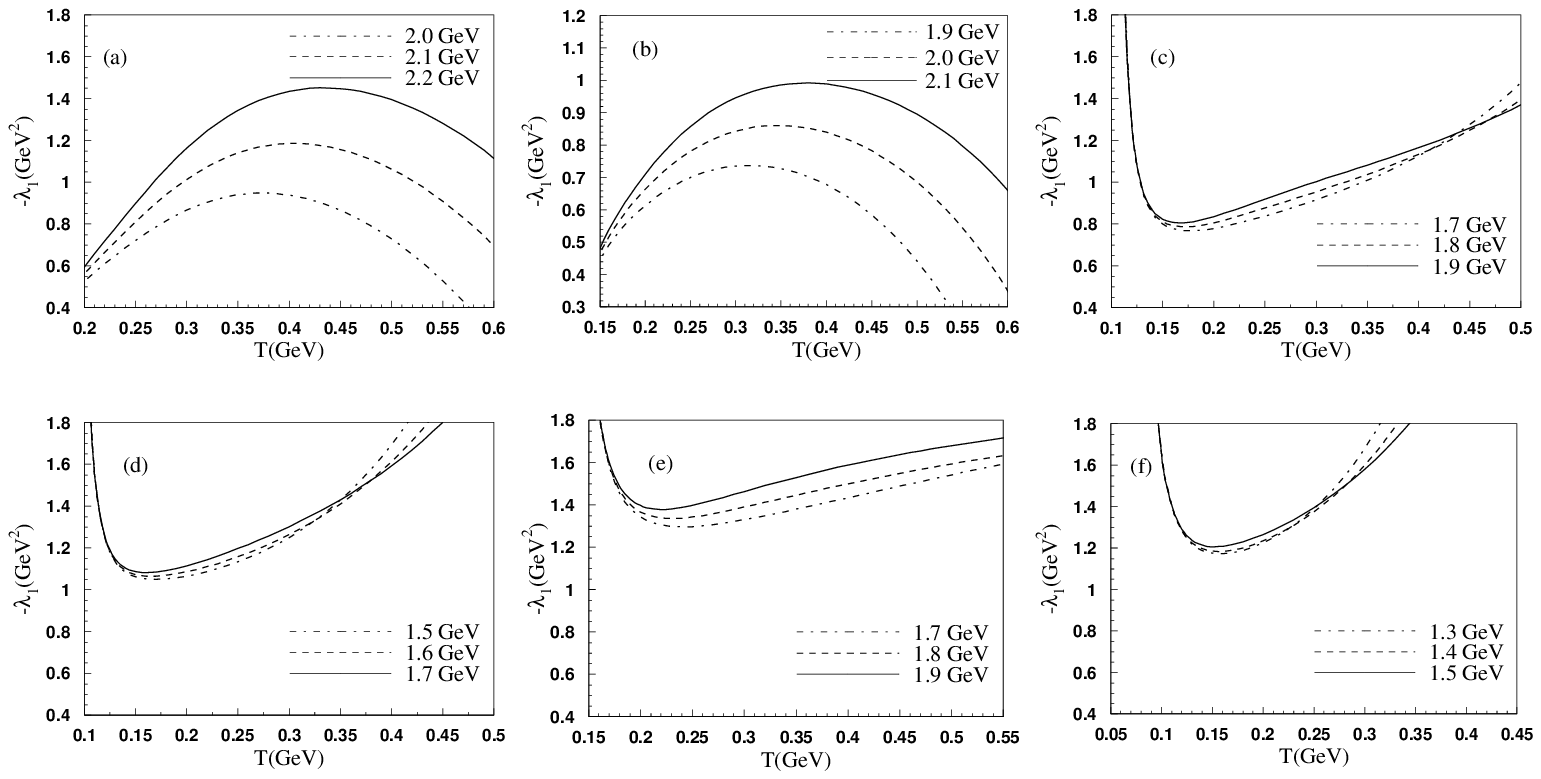}} \vfill \center{\mbox{Fig.
3.}}
\end{figure}

\begin{figure}
\epsfxsize=15cm \centerline{\epsffile{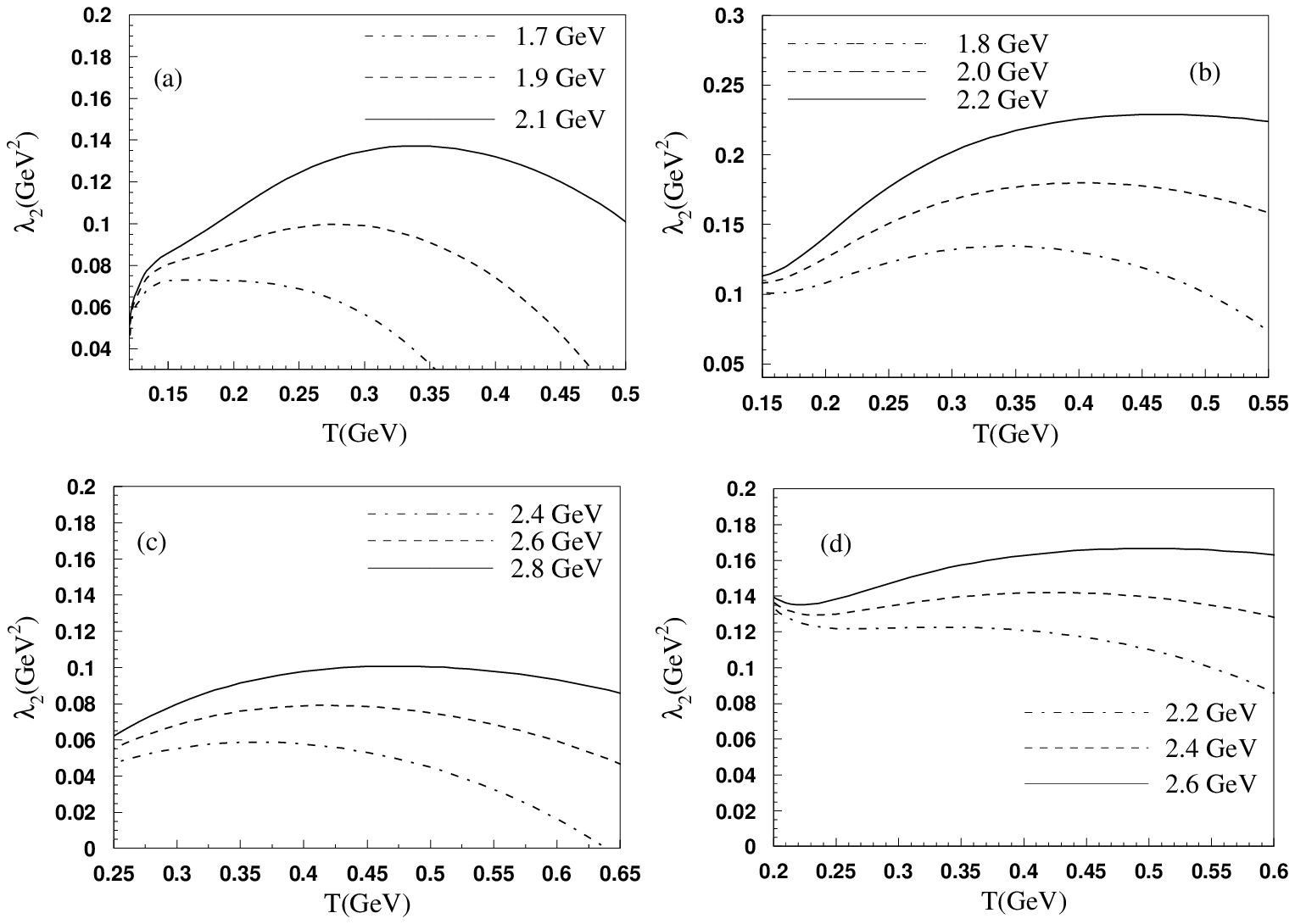}} \vfill \center{\mbox{Fig.
4.}}
\end{figure}
\end{document}